\documentclass[twocolumn,showpacs,preprintnumbers,amsmath,amssymb]{revtex4}

\usepackage{graphicx}% Include figure files
\usepackage{dcolumn}% Align table columns on decimal point
\usepackage{bm}% bold math
\usepackage{mathrsfs}
\usepackage{color}

\begin{document}

\title{
Alpha Clustering with a Hollow Structure \\
{\it --- Geometrical Structure of Alpha Clusters from Platonic Solids to Fullerene Shape}
}% 

\author{Akihiro Tohsaki$^1$ and Naoyuki Itagaki$^2$}

\affiliation{
$^1$Research Center for Nuclear Physics (RCNP), Osaka University,
10-1 Mihogaoka, Ibaraki, Osaka 567-0047, Japan
}

\affiliation{
$^2$Yukawa Institute for Theoretical Physics, Kyoto University,
Kitashirakawa Oiwake-Cho, Kyoto 606-8502, Japan
}

\date{\today}

\begin{abstract}
We study $\alpha$-cluster structure based on the geometric configurations 
with a microscopic framework, which takes full account of 
the Pauli principle, and which also employs an effective inter-nucleon force
including finite-range three-body terms
suitable for microscopic $\alpha$-cluster models. Here, special attention is focused 
upon the $\alpha$ clustering with a hollow structure;
all the $\alpha$ clusters are put on the surface of a sphere. 
All the Platonic solids (five regular polyhedra) and the fullerene-shaped 
polyhedron coming from icosahedral structure are considered. 
Furthermore, two configurations with dual polyhedra, 
hexahedron-octahedron and dodecahedron-icosahedron, 
are also scrutinized.
As a consequence, we insist on the possible 
existence of stable $\alpha$-clustering with a hollow structure for all the configurations.
Especially, two configurations, that is, dual polyhedra of dodecahedron-icosahedron and
fullerene, have a prominent hollow structure compared with other six configurations.
\end{abstract}

% PACS, the Physics and Astronomy
                             % Classification Scheme.
%\keywords{Suggested keywords}%Use showkeys class option if keyword
                              %display desired
\maketitle

%\section{Introduction}

Carbon atoms play an essential role 
in composing molecular structure related to geometric configuration 
in organic chemistry.  It is plausible that $\alpha$-particles 
are in the same situation in nuclear structure as carbon atoms 
in molecular structure because of their strong binding energy 
and the dual role of the Pauli principle. 
When two $\alpha$-particles are at a distance each other, 
the Pauli principle works attractively;
on the other hand, the strong repulsion acts on approaching two 
$\alpha$-particles. 
Not only cannot the $\alpha$ cluster easily break down
but also two $\alpha$ particles have a resonance state 
around the threshold energy. We can point out that three $\alpha$ clusters 
are loosely bound in making Borromean nucleus, which was predicted to be 
$\alpha$-cluster condensation~\cite{THSR}. 

Up to now, there have been many studies
on the geometrical structure of  $\alpha$ clusters based on 
the microscopic frameworks, which employ effective inter-nucleon forces 
and completely consider the Pauli principle simultaneously~\cite{Fujiwara}. 
Especially, the Brink-Bloch model is one of the suitable tools for 
studying the geometric structure of $\alpha$-clustering~\cite{Brink}. 
However, almost all the attempts have been restricted 
up to $4N$ nuclei of (0p)-shell region. Furthermore, the employed effective 
inter-nucleon force is different for every $\alpha$-type $4N$ nuclei 
because we do not have appropriate ones to reproduce 
the physical quantities in the wide mass number region from 
$\alpha$-particle to nuclear matter.  
For instance, the Volkov force~\cite{Volkov}, which is the most popular 
inter-nucleon force, includes the Majorana strength 
as an adjustable parameter for every nucleus. 
Nevertheless, the saturation property for the nuclear matter 
cannot be reproduced.         

Fortunately, the introduction of the finite-range three-body 
inter-nucleon force can elegantly overcome the defects of inter-nucleon force 
only with the two-body terms. The overall saturation property 
in the wide mass number region of $4N$ nuclei is well explained by using 
an effective inter-nucleon force with finite-range  three-body terms related 
to the density dependency.  The concepts of deciding the parameters 
in the effective inter-nucleon force are as follows: 
1) reasonable reproduction of the saturation property of $\alpha$, 
$^{16}$O, $^{40}$Ca and nuclear matter, 2) the reproduction of 
the phase shift of elastic $\alpha$-$\alpha$ scattering.  
In this report, we use Tohsaki F1 force~\cite{Tohsaki} (we call F1).  
Recent report by one of the authors verifies the validity of the F1 force 
for the unified understanding of the $^{12}$C and $^{16}$O~\cite{Itagaki-CO}. 
As a comparison, we show the results for the Brink-Boeker force~\cite{BB} 
(referred as BB-force), which has also no adjustable parameter, 
but which cannot reproduce the saturation property except for $\alpha$ 
particle and nuclear matter. 
Therefore, we think that F1 force is more reliable than BB-force, but
here we compare the results of two forces.

\begin{figure}[t]
        \centering
        \includegraphics[width=8.0cm]{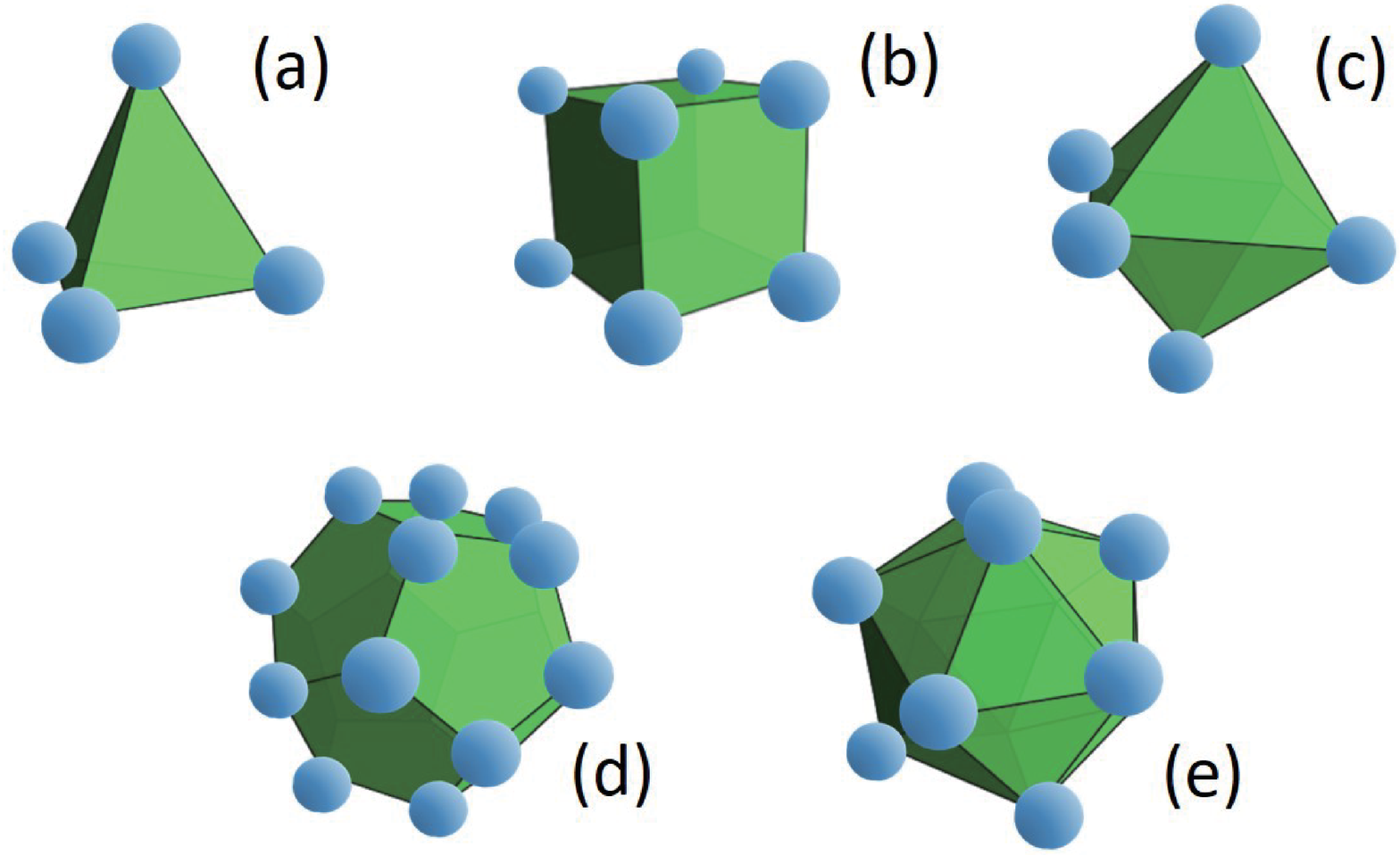}
        \includegraphics[width=6.0cm]{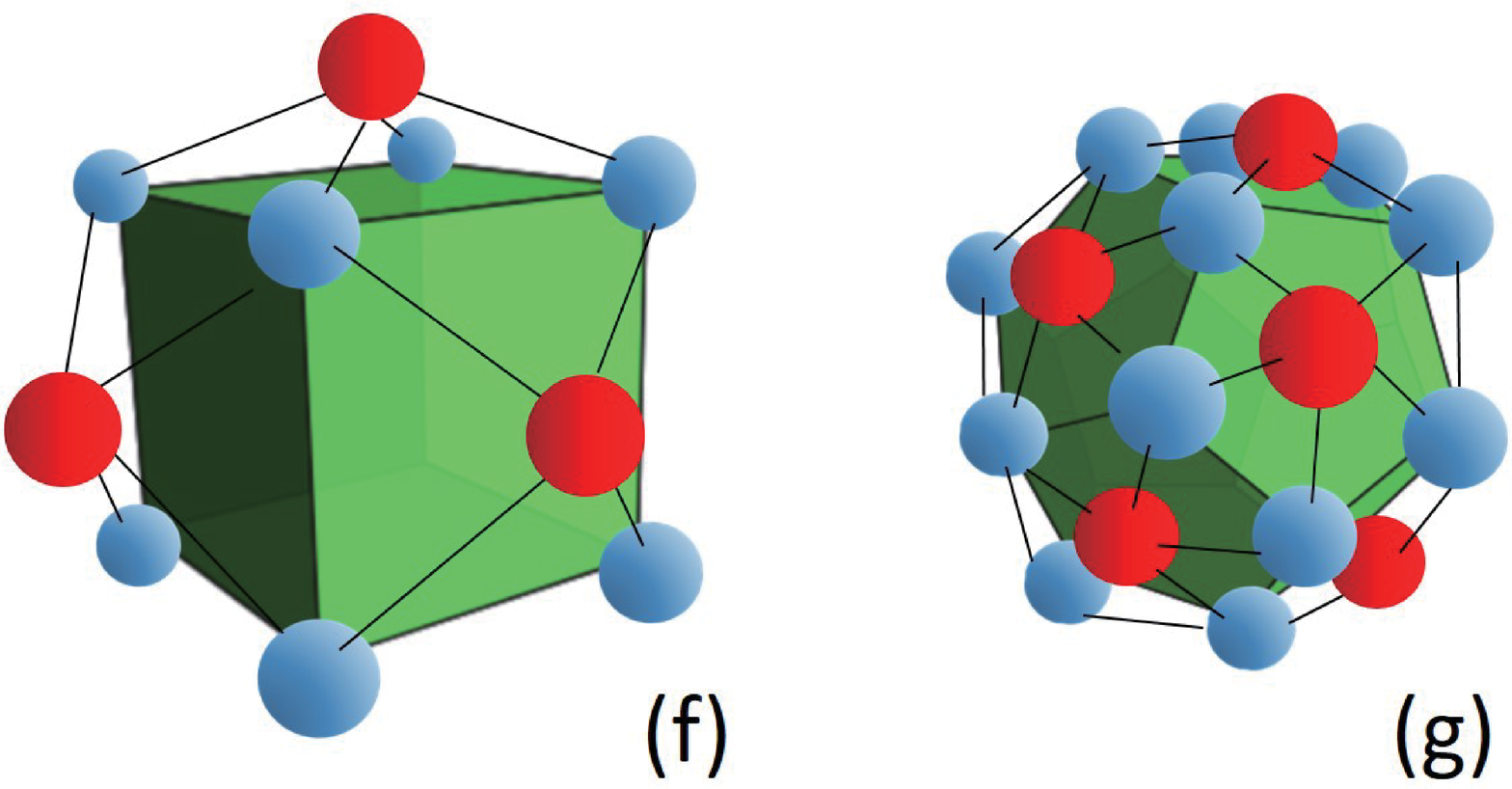}
        \includegraphics[width=6.0cm]{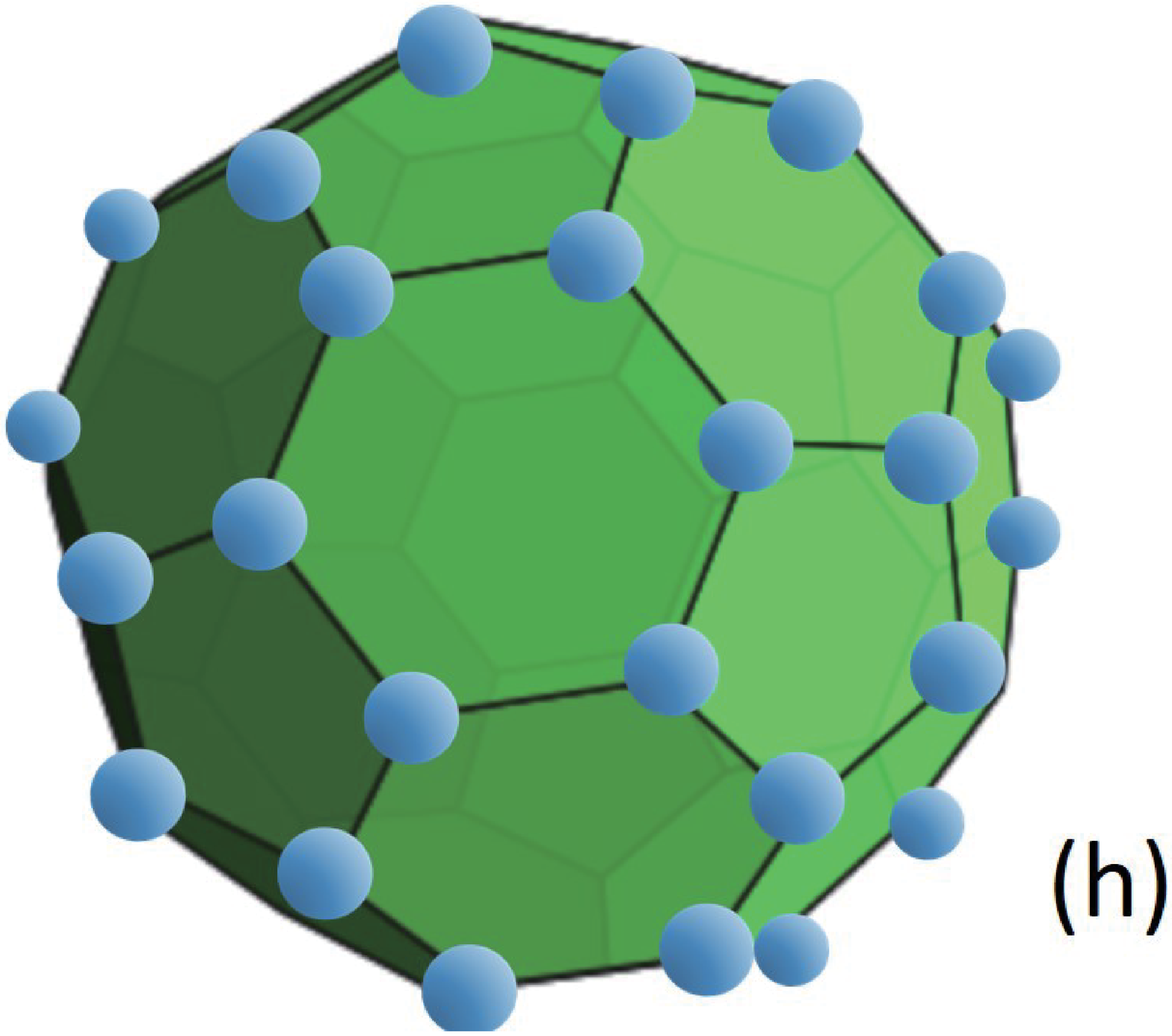}
        \caption{
Schematic figures for the prepared configurations,
where vertices on the polyhedra show the positions of the $\alpha$ clusters;
(a) tetrahedron, (b) hexahedron (cube), (c) octahedron, (d) dodecahedron,  (e) icosahedron,
(f) hexahedron-octahedron, (g) dodecahedron-icosahedron,
and (h) fullerene-shape polyhedron configurations. 
     }
\label{fullerene}
\end{figure}
As for the geometric configurations, first of all, we consider 
five Platonic solids (five regular polyhedra), 
of which vertices are positions of $\alpha$ clusters.
As schematically shown in Fig.~1, we prepare
(a) tetrahedron, (b) hexahedron (cube), (c) octahedron, (d) dodecahedron, and (e) icosahedron configurations.
They consists of 4, 8, 6, 20, and 12 $\alpha$ clusters corresponding to 
$^{16}$O, $^{32}$S, $^{24}$Mg, $^{80}$Zr, and $^{48}$Cr, respectively. 
In addition, we introduce their dual polyhedra, 
(f) hexahedron-octahedron and (g) dodecahedron-icosahedron;
the latter is related to the 
rhombic triacontahedron, which is the basic seed of quasi-crystal.
Here, hexahedron-octahedron is a combination of two Platonic solids,
hexahedron and octahedron. 
In Fig.~1 (f), blue balls are $\alpha$ clusters at the
vertices of hexahedron, and red balls are $\alpha$ clusters, which form pyramid shape together
with the four nearest (blue) $\alpha$ clusters. There are six red balls corresponding to the
number of faces of hexahedron, and these  six points form an octahedron shape.
In total, we have fourteen $\alpha$ clusters corresponding to $^{56}$Ni,
and the distances from the origin are taken to be common for all the fourteen $\alpha$'s.
Therefore, hexahedron and octahedron are inscribed in a common sphere.
If we start with an octahedron shape and add $\alpha$ clusters at the centers of the faces,
formation of completely the same solid is achieved, and 
hexahedron and octahedron are considered as a pair (dual polyhedra).
Another one is dodecahedron-icosahedron,
which is the combination of dodecahdron and icosahdron.
In Fig.~1 (g), twenty blue balls are $\alpha$ clusters at the
vertices of dodecahedron, and twelve red $\alpha$ clusters
are added at the center of twelve faces, which form icosahedron shape.
We have thirty two $\alpha$ clusters corresponding to $^{128}$Gd,
and the distances from the origin are taken to be common for all the $\alpha$'s;
dodecahedron and icosahedron are inscribed in a common sphere.
If we start with an icosahedron shape and add $\alpha$ clusters at the centers of the faces,
the same solid is formed, and 
dodecahedron and icosahedron are considered as a pair (dual polyhedra).
Note that tetrahedron is a self-dual polyhedron. 
If we apply the same procedure for the tetrahedron
(adding $\alpha$ clusters at the center of each surface and adjusting the distances
from the origin to be the same as those of $\alpha$'s at vertices), 
it becomes hexahedron (cube), which is already included in the model as (b).
Furthermore, we introduce (h) fullerene-shape polyhedron, which has sixty $\alpha$
clusters corresponding to $^{240}$120. 
The fullerene shaped nucleus, 
whose atomic number is 120, belongs to the ultra-super heavy 
region of nucleus. This is, of course, undiscovered until now.  
If the radius is very large, 
the configuration contains a big void inside of the sphere. 
In other words, we can imagine balloon-shaped nuclei
and takes the rhombic triacontahedron and the fullerene configurations 
as their plausible candidates. In order to study the stability of the structure 
in this report, we show an adiabatic-energy curve with respect to the radius 
of the sphere, which is the only variational parameter to see the property of 
the breathing mode. 
Carrying out the angular momentum projection is unnecessary because of the symmetric shape of the sphere.

% The paper is organized as follows. 
% The formulation is given in Sect.~\ref{model}. 
% In Sect.~\ref{results}, the AQCM results are shown. 
% Finally, in Sect.~\ref{summary} we summarize the results and give the main conclusions.

%\section{Model wave function and Hamiltonian \label{model}}

We employ Brink-Bloch type wave function, which takes complete account of the Pauli principle:
\begin{equation}
\Psi (\rho) = {\cal A} 
\{ 
\phi_1(\rho \bm R_1)
\phi_2(\rho \bm R_2) 
\cdot \cdot \cdot \cdot 
\phi_N(\rho \bm R_N) 
\}, 
\label{total-wf}
\end{equation} 
where ${\cal A}$ is the anti-symmetrization operator 
among all the nucleons. 
The $N\alpha$ clusters are on the surface of the sphere with the radius $\rho$ (fm), 
and the vectors $\bm R_1, \cdots, \bm R_N$ 
are the parameters on the dimensionless unit sphere.  
The $k$-th $\alpha$ cluster ($k = 1, 2, \cdot \cdot N$) wave function is written by
\begin{equation}
	\phi(\rho \bm R_k) =  \prod_{i,j=1,2}
	\left(\frac{1}{\pi b^2} \right)^{\frac{3}{4}}
	\exp \left[- {1 \over 2b^2} \left(\bm{r}^{ij}_k - \rho \bm{R}_k \right)^{2} \right] \chi^{ij}_k,
\label{Brink-wf}
\end{equation}
where $b$ is the nucleon size parameter, and $\chi^{ij}_k$ is a spin isospin wave function.
The vector $\bm{r}^{ij}_k$ is the real physical coordinate for the nucleon,
and $i$ and $j$ are labels for the spin and isospin, respectively, 
for the four nucleons in the $k$-th $\alpha$ clusters.
The four nucleons in the $k$-th $\alpha$ cluster share the common Gaussian center,
$\rho \bm{R}_k$.
We prepare eight sets of $\{ \bm R_1, \cdots, \bm R_N \}$
corresponding to the configurations in Fig.~1.
It is pointed out that all the configurations in Fig.~1 
inevitably contain their ground state components of harmonic oscillator 
wave functions at the small $\rho$ limit
owing to the anti-symmetrization effect. 
This is the same as the $\alpha$-condensation wave function (so-called THSR wave function), 
which includes the ground state component for every corresponding nucleus~\cite{Cond}.

The norm and energy kernel matrix elements
after carrying out the integration with respect to the real physical coordinates $\{ \bm{r}^{ij}_k \}$
are functions of variational parameter $\rho$:
$\langle \Psi (\rho') | \Psi (\rho) \rangle$
and
$\langle \Psi (\rho') |\hat{H}| \Psi (\rho) \rangle$,
where the Hamiltonian is given by
\begin{eqnarray}
\hat{H} = && -{\hbar^2 \over 2M}\sum_i \nabla^2_i - T_{cm}  \nonumber \\
&& +{1 \over 2!} \sum_{i,j} v^{(c)}_{ij} 
+{1 \over 2!} \sum_{i,j} v^{(2)}_{ij} 
+ {1 \over 3!} \sum_{i,j,k}  v^{(3)}_{ijk}.
\end{eqnarray}
The first and the second terms are the kinetic operator and the center of mass (c.m.) energy.
The third is the Coulomb operator running over the protons, 
and the fourth and the fifth terms are the effective inter-nucleon force 
separated by the two-body and three-body ones. 
The explicit form is written by the summation of Gaussian function:
\begin{eqnarray}
v^{(2)}_{ij} = 
\sum_{l=1}^3 && V^{(2)}_l 
 ((1-m^{(2)}_l) - m^{(2)}_l) P^{\sigma}_{ij} P^{\tau}_{ij} \nonumber \\
&& \times \exp[- (\bm r_i - \bm r_j )^2 / \beta_l^2]
\label{2body}
\end{eqnarray} 
and
\begin{eqnarray}
v^{(3)}_{ijk} = 
    \sum_{l=1}^3 && V^{(3)}_l 
\{ (1-m^{(3)}_l) - m^{(3)}_l \} P^{\sigma}_{ij} P^{\tau}_{ij} \nonumber \\
&& \times \{ (1-m^{(3)}_l) - m^{(3)}_l \} P^{\sigma}_{jk} P^{\tau}_{jk} \nonumber \\
&& \times \exp[- (\bm r_i - \bm r_j )^2 / \beta_l^2- (\bm r_j - \bm r_k )^2 / \beta_l^2],
\end{eqnarray}
where the exchange operators for the spin and isospin parts are expressed by $P^{\sigma}_{ij}$ and $P^{\tau}_{ij}$.
The force strengths for two- and three-body are written by 
$V_l^{(2)}$ and $V_l^{(3)}$, where their range parameters are given by $\beta_l$, and 
the Majorana strengths are $m_l^{(2)}$ and $m_l^{(3)}$.  
The force parameters of F1 are listed in Table I, where not only two-body 
but also three-body are given by the finite three-range description 
unlike the $\delta$-type zero-range force. The parameters for the range of the inter-nucleon 
force are taken to be the same in two-body and three-body parts. The BB force, 
which does not have the three-body terms, is listed in Table~II.

\begin{table} 
 \caption{ 
Parameter set for F1 interaction~\cite{Tohsaki}.
}

(a) two-body part\\
  \begin{tabular}{cccc} \hline \hline
 $l$  & $\beta_l$ (fm)  & $v^{(2)}_l$ (MeV) & $m^{(2)}_l$  \\ \hline
   1  &  2.5 & $-5.00$   & 0.75 \\  
   2  &  1.8 & $-43.51$  & 0.462  \\ 
   3  &  0.7 & $60.38$  & 0.522  \\ \hline 
  \end{tabular}   
\\

(b) three-body part \\
\begin{tabular}{cccc} \hline \hline
 $l$  &  $\beta_l$ (fm) & $v^{(3)}_l$ (MeV) & $m^{(3)}_l$  \\ \hline
   1  & 2.5 & $-0.31$ &  0.000 \\  
   2  &  1.8 & 7.73  & 0.000 \\  
   3  &   0.7 & 219.0 & 1.909 \\ \hline 
  \end{tabular}   
\end{table}

\begin{table} 
 \caption{ 
Parameter set for Brink-Boeker interaction~\cite{BB}.
}

  \begin{tabular}{cccc} \hline \hline
 $l$  & $\beta_l$ (fm)  & $v^{(2)}_l$ (MeV) & $m^{(2)}_l$  \\ \hline
   1  &  1.4 & $-140.6$   & 0.4864 \\  
   2  &  0.7 & $389.5$  & 0.5290  \\  \hline 
  \end{tabular}   

\end{table}

When all the position parameters $\{ \bm{R}_k \}$
are given a priori, after the integral with respect to the real physical coordinates $\{ \bm{r}^{ij}_k \}$, 
the Shr\"{o}dinger equation changes into Hill-Wheeler equation written by
\begin{equation}
\int 
\{
\langle \Psi (\rho') |\hat{H}| \Psi (\rho) \rangle
-E\langle \Psi (\rho') | \Psi (\rho) \rangle
\} f(\rho) d\rho
= 0
\end{equation}
where $f(\rho)$ is the weight function for the energy $E$. 
In this report, however, we focus upon the estimated adiabatic energy, 
\begin{equation}
E(\rho) =
{
\langle \Psi (\rho) |\hat{H}| \Psi (\rho) \rangle
\over 
\langle \Psi (\rho) | \Psi (\rho) \rangle,
} 
\end{equation}
to find out the bulk 
property of the geometric configuration of $\alpha$-clustering 
with a hollow structure.
The diagonal part of the norm kernel has the following property 
depending on the anti-symmetrization effect:
\begin{equation}
\lim_{\rho \to 0}  \langle \Psi (\rho) | \Psi (\rho) \rangle = 0
\end{equation}
and we define the normalization as
\begin{equation}
\lim_{\rho \to \infty}  \langle \Psi (\rho) | \Psi (\rho) \rangle = 1.
\end{equation}
Therefore, it is reasonable to define the Pauli index as
\begin{equation}
p_i(\rho) = 1- \langle \Psi (\rho) | \Psi (\rho) \rangle
\label{Pauli}
\end{equation}
for each configuration.

%\section{Results and discussion}

In Figs. 2 (for (a)-(e) in Fig.~1) 
and 3 (for (f)-(h) in Fig.~1), we show the adiabatic energy curves per $\alpha$ for the case of F1 force.  
The horizontal axis is the radius
$\rho$ in Eq.~\ref{total-wf}. 
The size parameter of single nucleon wave function $b$ 
is chosen to be 1.415 fm leading to the minimum of the binding energy of 
$\alpha$ particle, 27.500 MeV, reasonable comparing with the experimental value of
28.294 MeV; however here the basis of the energy is taken as the $N\alpha$ 
break-up energy. 
In all cases, the adiabatic energy curves have the energy pocket at shorter distances and barrier at larger distances.

\begin{figure}[t]
        \centering
        \includegraphics[width=6.0cm]{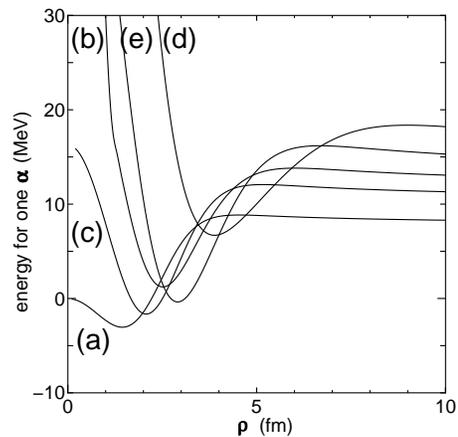}
        \caption{
Energy for one $\alpha$ 
as  a function of the radius ($\rho$ in the text), 
(a) tetrahedron, (b) hexahedron (cube), 
(c) octahedron, (d) dodecahedron, and (e) icosahedron. 
     }
\label{Platonic}
\end{figure}

\begin{figure}[t]
        \centering
        \includegraphics[width=6.0cm]{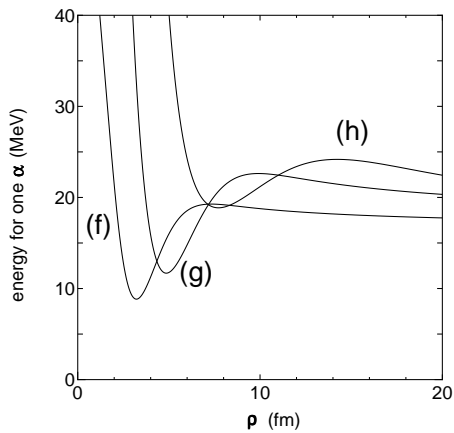}
        \caption{
Energy for one $\alpha$ 
as  a function of the radius ($\rho$ in the text), 
(f) 
hexahedron-octahedron, 
(g) dodecahedron-icosahedron, 
(h) fullerene.  
     }
\label{fullerene}
\end{figure}

\begin{table} 
 \caption{ 
 The physical quantities 
 at the (a) energy pocket and (b) barrier positions
 calculated using  F1 force.
 The radius ($\rho$ in the text), depth of the energy pocket and barrier, 
  $p_i(\rho)$ in Eq.~\ref{Pauli} are  listed. 
Here s.d. means the shortest distance of two $\alpha$ clusters 
at the fixed radius of $\rho$. 
}

(a) Pocket \\

  \begin{tabular}{lcccc} \hline \hline
      & $\rho$ (fm) & s.d.(fm) & depth (MeV)  & $p_i(\rho)$ \\ 
\hline
  P04(4)               & 1.4 &  2.29 & $-3.039$ & 0.994 \\
  P08(6)               & 2.1 &  2.97 & $-1.643$ & 0.986 \\
  P06(8)               & 2.5 &  2.89 & 1.228 & 0.998 \\
  P20(12)              & 2.9 & 3.05 & $-0.367$ & 1.000 \\
  P12(20)              & 3.9 & 2.78 & 6.700 & 1.000 \\
  P08-06(14)         & 3.2 & 2.94 & 8.838 & 1.000 \\
  P12-20(32)        & 4.9 & 3.14 & 11.694 & 1.000 \\
  Ful(60)              & 7.7 & 3.11 & 18.868 & 1.000 \\
  \hline
  \end{tabular}   

(b) Barrier

  \begin{tabular}{lcccc} \hline \hline
      & $\rho$ (fm) & s.d.(fm) & depth (MeV)  & $p_i(\rho)$ \\ 
\hline
  P04(4)               & 4.5 &  7.35 & 8.845 & 0.000 \\
  P08(6)               & 5.2 &  7.35 & 12.071 & 0.000 \\
  P06(8)               & 6.0 &  6.93 & 13.814 & 0.000 \\
  P20(12)              & 6.7 & 7.05 & 16.180 & 0.000 \\
  P12(20)              & 9.0 & 6.42 & 18.367 & 0.004 \\
  P08-06(14)         & 7.3 & 6.71 & 19.271 & 0.001 \\
  P12-20(32)        & 9.9 & 6.41 & 22.626 & 0.011 \\
  Ful(60)              & 14.2 & 5.73 & 24.180 & 0.094 \\
  \hline
  \end{tabular}   

\end{table}

\begin{table} 
 \caption{ 
  The physical quantities 
   at the (a) energy pocket and (b) barrier positions
   calculated using Brink-Boeker force.
 The radius ($\rho$ in the text), depth of the energy pocket and barrier, 
  $p_i(\rho)$ in Eq.~\ref{Pauli} are listed. 
Here s.d. means the shortest distance of two $\alpha$ clusters
at the fixed radius of $\rho$. 
}

(a) Pocket \\

  \begin{tabular}{lcccc} \hline \hline
      & $\rho$ (fm) & s.d.(fm) & depth (MeV)  &  $p_i(\rho)$ \\ 
\hline
  P04(4)               & 2.0 &  3.27 & 3.200 & 0.732 \\
  P08(6)               & 2.3 &  3.25 & 4.732 & 0.936 \\
  P06(8)               & 2.8 &  3.23 & 8.458 & 0.972 \\
  P20(12)              & 3.1 & 3.26 & 7.841 & 0.999 \\
  P12(20)              & 4.9 & 3.50 & 15.803 & 0.997 \\
  P08-06(14)         & 3.6 & 3.31 & 16.079 & 0.998 \\
  P12-20(32)         & 5.3 & 3.40 & 18.540 & 1.000 \\
  Ful(60)              & 10.1 & 4.08 & 24.807 & 0.996 \\
  \hline
  \end{tabular}   

(b) Barrier

  \begin{tabular}{lcccc} \hline \hline
      & $\rho$ (fm) & s.d.(fm) & depth (MeV)  &  $p_i(\rho)$ \\ 
\hline
  P04(4)               & 4.2 &  6.86 & 9.017 & 0.000 \\
  P08(6)               & 4.7 &  6.65 & 12.325 & 0.000 \\
  P06(8)               & 5.5 &  6.35 & 14.117 & 0.002 \\
  P20(12)              & 6.1 & 6.41 & 16.605 & 0.004 \\
  P12(20)              & 8.2 & 5.85 & 18.867 & 0.021 \\
  P08-06(14)          & 6.7 & 6.16 & 20.102 & 0.007 \\
  P12-20(32)         & 9.1 & 5.83 & 23.295 & 0.047 \\
  Ful(60)              & 12.7 & 5.13 & 25.189 & 0.384 \\
  \hline
  \end{tabular}   

\end{table}

In Table III, the physical quantities at these  (a) energy pocket and (b) barrier positions
in the case of  F1 force are listed.
Here, the acronym s.d. means the shortest distance of two $\alpha$ clusters 
at the fixed radius of $\rho$, and
P04(4), P08(6), P06(8), P20(12), P08-06(14), P12-20(32), and Ful(60) are
configurations (a)-(h) in Fig.~1,
where the values in the parentheses show the numbers of $\alpha$ clusters.
The characteristic features are the following;
1) there are stable energy pockets in all the cases,
2) the energy pocket is protected by the competition of the Coulomb repulsion 
and the Pauli principle,
3) the values of s.d. are almost the same for all the cases of energy pocket and barrier, 
but the rhombic triacontahedron and the fullerene shape have comparably 
large distance of two $\alpha$ particles,
4) the Pauli index becomes almost 1 around the energy pocket, 
and the increase starts at the barrier position where the index is 0.
All the quantities inevitably include the spurious c.m. 
energy of each $\alpha$ particle to be removed due to the adiabatic treatment. 
The removal procedure is not so easy but possible. 
We think, however, that it is useful to see the general trends of the 
configuration via the adiabatic energy curves. 

Especially, the fullerene shape can exist stably, 
but its radius is very large, around 7.7 fm. 
We predict a big void surrounded by 60 $\alpha$ particles,
but we see that the gap of the barrier height and the pocket decreases 
when the number of $\alpha$ particles increases. 
Note again that every shape of geometric configuration inevitably includes 
the ground state component for every corresponding nucleus. 
Therefore, it is desired to quantitatively scrutinize the stability of 
the balloon structure of $\alpha$ particles by using more reliable 
wave function.  One of the candidates of the wave function is to employ 
the THSR ansatz, which reasonably describes the $\alpha$ condensation. 

We show the same quantities for the case of BB force in Table IV. 
The $b$ parameter is taken to be 1.409 fm which gives the binding energy of 
27.375 MeV for $\alpha$ particle. Surprisingly enough, even fullerene 
shaped $\alpha$ clustering has an energy pocket; however slightly shallower than the case 
of F1 force. On the other hand, the barrier position shifts inside. 
The general trends do not change so much comparing with the case of F1 force, 
but physical quantities largely change. For instance, the tetrahedron shape 
correctly represents the ground state of $^{16}$O, 
which is well reproduced by the F1 force, 
but this configuration gives very much underbinding in the BB force case; 
the energy shows the value of unbound region.

%\section{Concluding remark}
In this report, full microscopic calculations are carried out for the balloon shaped
$\alpha$ clustering; typical eight examples, five Platonic solids,
two cases of dual polyhedra, and fullerene shape are examined. 
All the configurations have the energy pocket with respect to the balloon 
radius in the F1 force cases and even in the BB force cases. 
It is pointed out that an exotic nucleus beyond super-heavy region, 
that is, fullerene shaped configuration of α clusters is stable. 
Here we avoid studying individual nuclei, 
because it is necessary to dynamically consider each nucleus, and this is
another task. Namely, the next step of this investigation is to come into 
the clarification of the properties on individual nuclei.   
Is it crazy to imagine balloon nuclei which consist of $\alpha$-clusters? 
Our answer is `No'.  Reliable effective inter-nucleon force and the complete 
consideration of the Pauli principle make it possible to give a correct answer.
Surprisingly enough, the present report is to predict that even ultra-super 
heavy nuclei can exist in such a presence form as a geometric structure 
with a void, which requires us to revise the empirical formula for the nuclear radius.

\begin{acknowledgments}
One of the authors (A.T) has discussed the fullerene shaped $\alpha$-clusters 
with the late W. Greiner, for whom the authors are now grateful. We also thank H. Horiuchi, 
Y. Funaki, P. Schuck and G. R\"{o}pke for their fruitful discussions. 
Numerical calculation has been performed at Yukawa Institute for Theoretical Physics, Kyoto University. This work was supported by     
JSPS KAKENHI Grant Number 17K05440.
\end{acknowledgments}

\end{document}